\documentclass[11pt,twoside]{article}
\usepackage{asp2010}

\resetcounters

\begin{document}

\title{The local LIRG NGC\,5135: AGN and SN activity traced by NIR IFU spectroscopy}
\author{Alejandro G.~Bedregal$^{1,2}$, Luis Colina$^2$, Ruyman Azzollini$^2$, Santiago Arribas$^2$, Almudena Alonso-Herrero$^2$}

\affil{$^1$Departamento de Astrof\'{\i}sica, Facultad de Ciencias F\'isicas, Universidad Complutense de Madrid, 28040 Madrid, Spain}
\affil{$^2$Departamento de Astrof\'{\i}sica, Centro de Astrobiolog\'ia (CSIC/INTA), Instituto Nacional de T\'ecnica Aeroespacial, Crta.~de Torrej\'on a Ajalvir, km 4, 28850 Torrej\'on de Ardoz, Madrid, Spain}

\begin{abstract}
By observing the local luminous infrared galaxy NGC\,5135 with the near-IR IFU spectrograph SINFONI (ESO VLT), we report a $\rm \sim600\,pc$ (in projection) AGN outflow traced by [SiVI]${\rm \lambda1.96\,\mu m}$ emission. This is the largest outflow traced by a coronal line ever reported. Its large spatial scale suggests that shocks, in addition to AGN continuum emission, are needed to locally produce [SiVI] emission. We also show, for the first time, clear kinematical evidence of the AGN-outflow vs.~ISM interaction through variations in the 2D velocity fields of different gas phases. Such local perturbations in the kinematics clearly match the outflow structure. 

We use the [FeII]${\rm \lambda1.64\,\mu m}$ emission, a supernovae tracer, to estimate the supernovae rate in different star-forming knots ($\rm \sim 250\,pc$ across) within the central 2.3\,kpc of NGC\,5135. The estimated supernovae rates go from $\rm 0.02-0.08\,yr^{-1}$ being in excellent agreement with predictions from $\rm 6\,cm$ radio emission in the same areas.

\end{abstract}

\section{Introduction}
Since their discovery \citep{A, B}, the importance of low-redshift ($z$) Luminous ($10^{11}L_\odot\le L_{IR}\le 10^{12}L_\odot$, LIRG) and Ultraluminous
($L_{IR}\ge10^{12}L_\odot$, ULIRG) Infrared Galaxies has been widely recognized. While LIRGs appear to be mostly spirals \citep{C, D},
ULIRGs are strongly interacting systems and mergers \citep[e.g.][]{E} evolving into intermediate-mass
ellipticals \citep[e.g.][]{F}. 
Local (U)LIRGs have been proposed as
possible counterparts of the submillimeter population observed at higher
$z$ \citep[for a review]{G}. Also, cosmological surveys with {\it Spitzer}
have shown that the majority of infrared (IR) selected galaxies at $z\le1$ are in the LIRG
class, while LIRGs and ULIRGs make a significant contribution to the IR
galaxy population and to the star formation at $1<z<2$ and $z\ge2$, respectively  \citep[e.g.][]{H, I}.

Detailed investigations of the physical properties, stellar populations, AGN-starburst connection and
gas flows on these complex systems can only be obtained through integral field spectroscopy (IFS). Initial
studies of small samples of (U)LIRGs based on 4-meter class telescope optical IFS have already been obtained \citep[e.g.][]{J, K, L}. 
To extend these studies to larger samples, and also to the near-IR, we have started a survey of
 low-$z$ (U)LIRGs using state-of-the-art IFS
like VLT/VIMOS \citep[optical:][]{M, N} and VLT/SINFONI \citep[near-IR:][]{O, P}. This survey
 will allow us to characterize the kpc-scale ionization and kinematics
of a representative sample of low-$z$ (U)LIRGs covering a wide luminosity range, several morphologies from spirals
to interacting and advanced mergers, as well as different classes of activity. This will also form a local reference for
future IFS studies of high-$z$ IR galaxies with instruments such as the {\it Near-IR Spectrograph} ({\it NIRSpec}) and 
{\it Mid-IR Instrument} ({\it MIRI}) on board of the {\it James Webb Space Telescope} (Gardner et al.\ 2006).

As part of our survey, we present some of our first results with SINFONI showing the power
of near-IR IFS by
studying the local LIRG NGC\,5135. This is an SBab galaxy at
$z=0.01396$ (from {\tt NED}\footnote{http://nedwww.ipac.caltech.edu/}, at $\rm \approx
58.7\,Mpc$ assuming $H_0=70\,\rm km\,s^{-1}\,Mpc^{-1}$) which belongs to a group
of seven galaxies \citep{R}. Its dual nature as a
starburst hosting an AGN and its almost face-on sky orientation make NGC\,5135
an ideal prototype-object for detailed studies of these hybrid systems. In this proceeding, we focus on the AGN and SN activity of this galaxy, stressing their interaction with the local interstellar medium (ISM).

\section{AGN-outflow and its interaction with the local ISM}

The ionization potential required to produce the [SiVI]$\lambda 1.96 \mu$m line
(167\,eV) is usually associated to Seyfert activity where the gas is
excited just outside the broad line regions of AGNs \citep[e.g.][]{S}. 

\begin{figure*}
\includegraphics[scale=0.32]{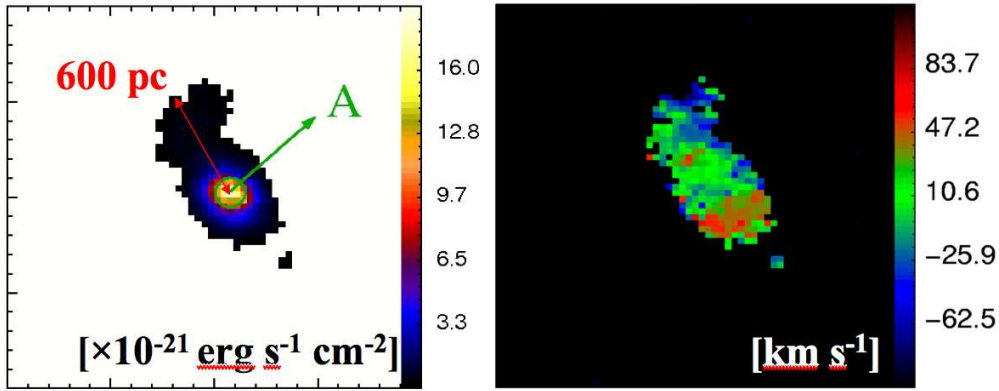}
\caption{\label{fig:Si6_FluxVel}\small [SiVI] flux and velocity maps within the central 2.3\,kpc field-of-view of NGC\,5135. The AGN is in region {\it A}. North is up, East is left.}
\end{figure*}

 As we can see in the Fig.\,\ref{fig:Si6_FluxVel} flux map, the [SiVI] line traces the galaxy nucleus and it also presents a weaker ``plume'' to the North-East (NE).
This particular region, in terms of projected spatial scales, is the largest reported in literature for a coronal line ($\rm \approx 600\,pc$ in this case) being $\sim 4\,\times$ larger than previous reports on
different Seyfert galaxy samples \citep[e.g.][]{T, S}. This finding, together with evidence for a similar (weaker) structure to the South-West (SW), suggest that we have detected, for the first time, the presence of ionizing cones in NGC\,5135. According to
\citet{S}, the morphology of [SiVI] and other coronal
gas is preferably aligned with the direction of the traditional
lower-ionization cones (i.\,e.\ traced by [OIII]) seen in Seyfert galaxies.

\begin{figure}
\includegraphics[scale=0.35]{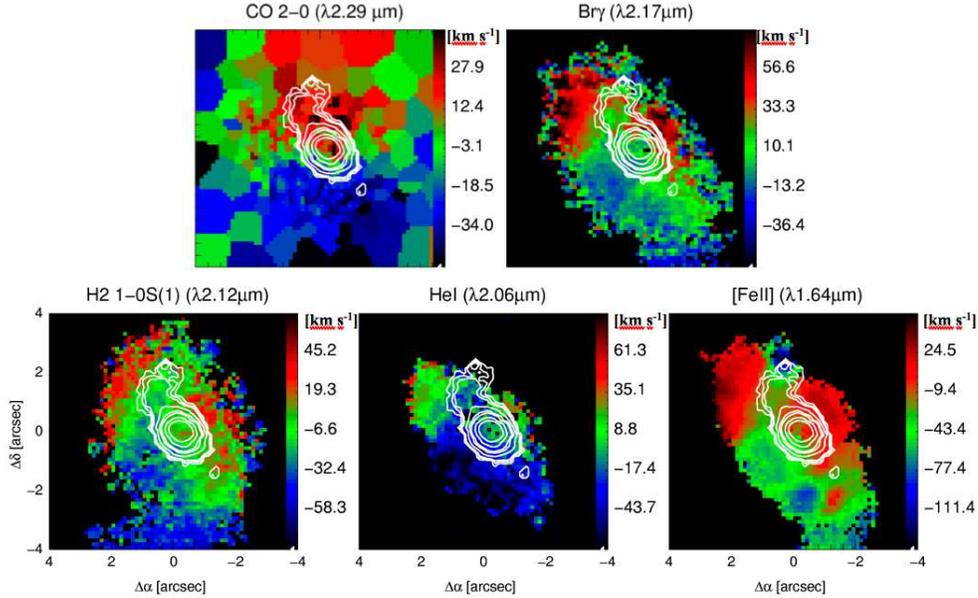}
\caption{\label{fig:Velmaps_Si6flx}\small Stellar and gas velocity fields with [SiVI] emission contours overplotted. Clearly the [SiVI] outflow (NE) matches with blueshifted regions in all gas phases.}
\end{figure}

Now we compare the information from [SiVI] with the kinematics of the other galaxy components to disentangle the real influence of the AGN on its neighborhood. A key observation comes from Fig.\,\ref{fig:Velmaps_Si6flx}. We plot the velocity fields of CO, Br$\gamma$, H$_2$, HeI and [FeII] while overplotting the flux contours of [SiVI] emission. In most of the panels we clearly see how the AGN NE outflow coincides with gas blueshifted regions in the North half of our field-of-view (FoV). 

As we see, the stellar and many of the gas velocity ($V$) fields ruffly coincide in the sense of being redshifted to the North and blueshifted to the South of our FoV. Differences, however, appear between stellar and gas components in the redshifted (North) region, where gas phases present a "double lobe'' structure in $V$, totally absent in the stars. The spatial coincidence of the [SiVI] 'plume' structure with this area suggests an interaction between  the AGN-outflow and the surrounding gas, blueshifting their $V$ fields along the projected cone.

\section{[FeII] as a Supernovae tracer}
For many years different authors have suggested that the [FeII]-emission from
galaxies traces the fast shocks produced by supernovae (SN) remnants and so, their SN activity
\citep[e.g.][]{U, V, W}. 

\begin{figure}
\includegraphics[scale=0.48]{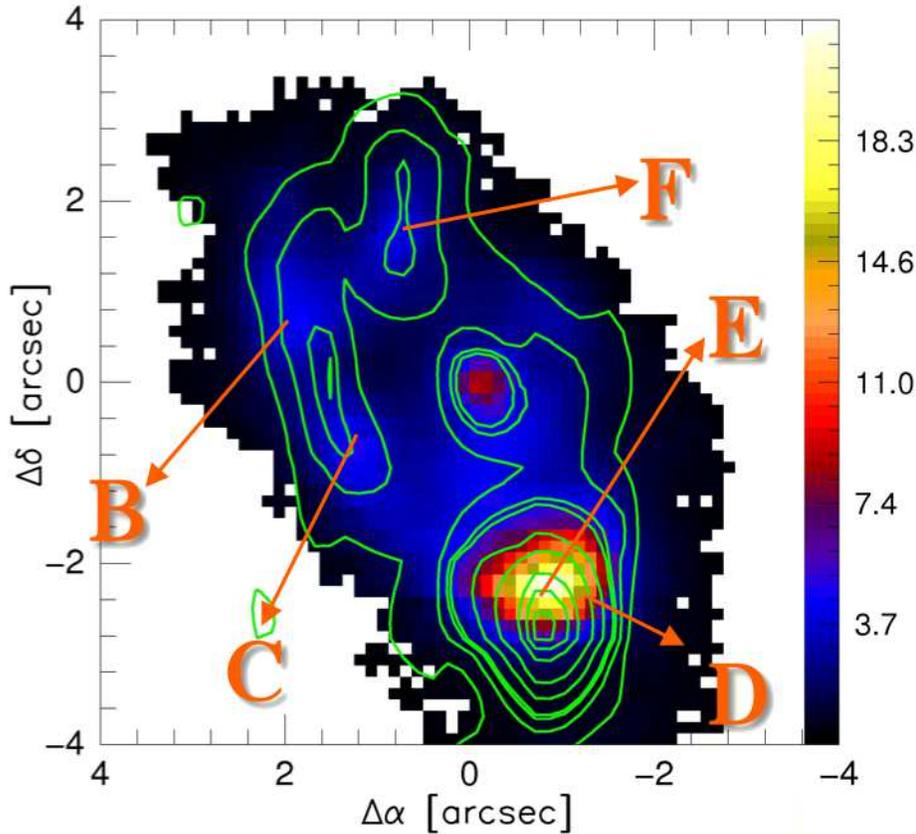}
\caption{\label{fig:feII_radio}\small [FeII] flux map of the central 2.3\,kpc of NGC\,5135. The
  contours are the $\rm 6\,cm$ VLA radio emission from Ulvestad \& Wilson (1989).}
\end{figure}

Different authors have derived empirical and theoretical relations we can use to
estimate the SN rate directly from the [FeII] emission \citep{X, V}
and from 6\,cm radio data \citep{Y, Z, A2}. 
We select five regions (labeled from {\it B}-{\it F} in Fig.\,\ref{fig:feII_radio}) corresponding to Br$\gamma$ and [FeII] peaks (starforming regions) using circular apertures of $\rm 0.91^{\prime \prime}$
($\rm \approx260\,pc$ in diameter). 

In Table\,\ref{tab:sne_rate_prime} we present the SN rates derived from [FeII] and 6\,cm radio data \citep{B2} for regions {\it B}-{\it F}.

\begin{table*}  
\begin{center}
 \caption{SN Rates derived from [FeII] and Radio emission \label{tab:sne_rate_prime}} 
 \begin{tabular}{@{}|l|cc||ccc|@{}}
\hline
 \scriptsize{Reg$^a$}        &  \scriptsize{Ca97$_{\rm [FeII]}$} & \scriptsize{A-H03$_{\rm [FeII]}$} & \scriptsize{Hu94$_{\rm 6cm}$} &
  \scriptsize{Ne04$_{\rm 6cm}$} & \scriptsize{P-O95$_{\rm 6cm}^b$}\\
                         & $\rm [yr^{-1}]$   & $\rm [yr^{-1}]$ & $\rm [yr^{-1}]$ & $\rm [yr^{-1}]$ & $\rm [yr^{-1}]$\\
\hline
{\it B}     & 0.012 $\pm$ 0.002&  0.017 $\pm$ 0.004 & 0.012 $\pm$ 0.004  & 0.019 $\pm$ 0.007 & 0.013-0.014 \\[0.1cm]
{\it C}     & 0.012 $\pm$ 0.003&  0.018 $\pm$ 0.004 & 0.010 $\pm$ 0.005  & 0.016 $\pm$ 0.008 & 0.011-0.012 \\[0.1cm]
{\it D}     & 0.024 $\pm$ 0.005&  0.036 $\pm$ 0.008 & 0.032 $\pm$ 0.075  & 0.051 $\pm$ 0.118 & 0.036-0.037 \\[0.1cm]
{\it E}     & 0.051 $\pm$ 0.012&  0.076 $\pm$ 0.016 & 0.064 $\pm$ 0.001  & 0.100 $\pm$ 0.002 & 0.071-0.074 \\[0.1cm]
{\it F}     & 0.011 $\pm$ 0.002&  0.016 $\pm$ 0.004 & 0.012 $\pm$ 0.004  & 0.019 $\pm$ 0.006 & 0.018-0.019 \\[0.1cm]
\hline
\end{tabular}\\
\end{center}
\footnotesize{($^a$) {\it B}-{\it F} regions with aperture of $\rm 0.91^{\prime \prime}$.}
\footnotesize{($^b$) For \citet{Z} models, we present ranges of SN
  rates for each region, where different upper- and lower-limit
  masses have been considered for a Salpeter IMF.}
\end{table*}

Table\,\ref{tab:sne_rate_prime} clearly shows the excellent agreement between [FeII] and radio SN rate predictions for these $\rm \sim 250\,pc$-scale regions. This agreement is totally independent of the correlation or model used to estimate the SN rate. These results provide additional support to previous findings in normal starburst galaxies like M82 and NGC\,253 where similar SN rates are predicted from [FeII] and $\rm 6\,cm$ radio emission. As a comparison, the predictions in this table are one order of magnitude lower than those of the brightest radio region in Arp\,299 \citep[$0.5$-$1.0\, \rm yr^{-1}$,][]{A2}, a bright IR luminous galaxy.

\section{Conclusions}

Our main conclusions are

\begin{itemize}

\item We report a $\rm \sim600\,pc$ (in projection) AGN outflow traced by [SiVI] emission. This is the largest outflow traced by a coronal line ever reported. This structure is at
least $4 \times$ larger than any previous detection in active
galaxies. Pointing in opposite direction, a fainter counter-[SiVI]-cone has been also
detected. 

\item Using 2D kinematics, we have found clear evidence of AGN-outflow versus ISM interaction. The different gas phases show perturbed velocity fields along the AGN outflow structure.

\item The SN rates derived from the [FeII] emission are in excellent agreement with
$\rm 6\,cm$ radio emission predictions, reinforcing the use of [FeII] as a SN
activity tracer. Typical rates between $0.02$-$\rm 0.08\,yr^{-1}$ were found for
individual $\rm \sim250\,pc$ regions. 

\end{itemize}

\acknowledgements  Based on observations carried out at the European Southern Observatory,
Paranal (Chile), program 077.B-0151(A). This work has been supported by the
Spanish Ministry for Science and Innovation under grant ESP2007-65475-C02-01. A.G.B. has been supported by the Programa Nacional de Astronom\'{i}a y Astrof\'{i}sica of the Spanish Ministry of Science and Innovation under grant AYA2007-67752-C03-03.

\bibliography{bedregal_guilin2010.bib}

\begin{thebibliography}{}
\expandafter\ifx\csname natexlab\endcsname\relax\def\natexlab#1{#1}\fi
\expandafter\ifx\csname url\endcsname\relax
  \def\url#1{\texttt{#1}}\fi
\expandafter\ifx\csname urlprefix\endcsname\relax\def\urlprefix{URL }\fi
\providecommand{\eprint}[2][]{\url{#2}}

\bibitem[{Alonso-Herrero et~al.(2009)Alonso-Herrero, Garc\'ia-Mar\'in,
  Monreal-Ibero, Colina, Arribas, Alfonso-Garz\'on, \& Labiano}]{L}
Alonso-Herrero, A., Garc\'ia-Mar\'in, M., Monreal-Ibero, A., Colina, L.,
  Arribas, S., Alfonso-Garz\'on, J., \& Labiano, A. 2009, A\&A, 506, 1541

\bibitem[{Alonso-Herrero et~al.(2006)Alonso-Herrero, Rieke, Rieke, Colina,
  P\'{e}rez-Gonz\'{a}lez, \& Ryder}]{D}
Alonso-Herrero, A., Rieke, G.~H., Rieke, M.~J., Colina, L.,
  P\'{e}rez-Gonz\'{a}lez, P.~G., \& Ryder, S.~D. 2006, ApJ, 650, 835

\bibitem[{Alonso-Herrero et~al.(2003)Alonso-Herrero, Rieke, Rieke, \&
  Kelly}]{V}
Alonso-Herrero, A., Rieke, G.~H., Rieke, M.~J., \& Kelly, D.~M. 2003, AJ, 125,
  1210

\bibitem[{Arribas et~al.(2004)Arribas, Bushouse, Lucas, Colina, \& Borne}]{C}
Arribas, S., Bushouse, H., Lucas, R.~A., Colina, L., \& Borne, K.~D. 2004, AJ,
  127, 2522

\bibitem[{Arribas et~al.(2008)Arribas, Colina, Monreal-Ibero, Alfonso,
  Garc\'ia-Mar\'in, \& Alonso-Herrero}]{M}
Arribas, S., Colina, L., Monreal-Ibero, A., Alfonso, J., Garc\'ia-Mar\'in, M.,
  \& Alonso-Herrero, A. 2008, A\&A, 479, 687

\bibitem[{Bedregal et~al.(2009)Bedregal, Colina, Alonso-Herrero, \&
  Arribas}]{O}
Bedregal, A.~G., Colina, L., Alonso-Herrero, A., \& Arribas, S. 2009, ApJ, 698,
  1852

\bibitem[{Bedregal et~al.(2011)Bedregal, Colina, Azzollini, Arribas, \&
  Alonso-Herrero}]{P}
Bedregal, A.~G., Colina, L., Azzollini, R., Arribas, S., \& Alonso-Herrero, A.
  2011, submitted to ApJ

\bibitem[{Blain et~al.(2002)Blain, Smail, Ivison, Kneib, \& Frayer}]{G}
Blain, A.~W., Smail, I., Ivison, R.~J., Kneib, J.-P., \& Frayer, D.~T. 2002,
  PhR, 369, 111

\bibitem[{Bushouse et~al.(2002)Bushouse, Borne, Colina, Lucas, Rowan-Robinson,
  Baker, Clements, Lawrence, \& Oliver}]{E}
Bushouse, H.~A., Borne, K.~D., Colina, L., Lucas, R.~A., Rowan-Robinson, M.,
  Baker, A.~C., Clements, D.~L., Lawrence, A., \& Oliver, S. 2002, ApJS, 138, 1

\bibitem[{Calzetti(1997)}]{X}
Calzetti, D. 1997, AJ, 113, 1

\bibitem[{Caputi et~al.(2007)Caputi, Lagache, Yan, Dole, Bavouzet, Le~Floc'h,
  Choi, Helou, \& Reddy}]{I}
Caputi, K.~I., Lagache, G., Yan, L., Dole, H., Bavouzet, N., Le~Floc'h, E.,
  Choi, P.~I., Helou, G., \& Reddy, N. 2007, ApJ, 660, 97

\bibitem[{Colina(1993)}]{U}
Colina, L. 1993, ApJ, 411, 565

\bibitem[{Colina et~al.(2005)Colina, Arribas, \& Monreal-Ibero}]{J}
Colina, L., Arribas, S., \& Monreal-Ibero, A. 2005, ApJ, 621, 725

\bibitem[{Genzel et~al.(2001)Genzel, Tacconi, Rigopoulou, Lutz, \& Tecza}]{F}
Genzel, R., Tacconi, L.~J., Rigopoulou, D., Lutz, D., \& Tecza, M. 2001, ApJ,
  563, 527

\bibitem[{Huang et~al.(1994)Huang, Thuan, Chevalier, Condon, \& Yin}]{Y}
Huang, Z.~P., Thuan, T.~X., Chevalier, R.~A., Condon, J.~J., \& Yin, Q.~F.
  1994, ApJ, 424, 114

\bibitem[{Kleinmann \& Low(1970)}]{A}
Kleinmann, D.~E., \& Low, F.~J. 1970, ApJ, 159, 165

\bibitem[{Kollatschny \& Fricke(1989)}]{R}
Kollatschny, W., \& Fricke, K.~J. 1989, A\&A, 219, 34

\bibitem[{Labrie \& Pritchet(2006)}]{W}
Labrie, K., \& Pritchet, C.~J. 2006, ApJS, 166, 188

\bibitem[{Monreal-Ibero et~al.(2006)Monreal-Ibero, Arribas, \& Colina}]{K}
Monreal-Ibero, A., Arribas, S., \& Colina, L. 2006, ApJ, 637, 138

\bibitem[{Neff et~al.(2004)Neff, Ulvestad, \& Teng}]{A2}
Neff, S.~G., Ulvestad, J.~S., \& Teng, S.~H. 2004, ApJ, 611, 186

\bibitem[{P\'erez-Gonz\'alez et~al.(2005)P\'erez-Gonz\'alez, Rieke, Egami, \&
  et~al.}]{H}
P\'erez-Gonz\'alez, P., Rieke, G.~H., Egami, E., \& et~al. 2005, ApJ, 630, 82

\bibitem[{P\'erez-Olea \& Colina(1995)}]{Z}
P\'erez-Olea, D.~E., \& Colina, L. 1995, MNRAS, 277, 857

\bibitem[{Prieto et~al.(2005)Prieto, Marco, \& Gallimore}]{T}
Prieto, M.~A., Marco, O., \& Gallimore, J. 2005, MNRAS, 364, 28

\bibitem[{Rieke \& Low(1972)}]{B}
Rieke, G.~H., \& Low, F.~J. 1972, ApJ, 176, 95

\bibitem[{Rodr\'iguez-Ardila et~al.(2006)Rodr\'iguez-Ardila, Prieto, Viegas, \&
  Gruenwald}]{S}
Rodr\'iguez-Ardila, A., Prieto, M.~A., Viegas, S., \& Gruenwald, R. 2006, ApJ,
  653, 1098

\bibitem[{Rodr\'iguez-Zaur\'in et~al.(2010)Rodr\'iguez-Zaur\'in, Arribas,
  Monreal-Iber\'o, Colina, Alonso-Herrero, \& Alfonso-Garz\'on}]{N}
Rodr\'iguez-Zaur\'in, J., Arribas, S., Monreal-Iber\'o, A., Colina, L.,
  Alonso-Herrero, A., \& Alfonso-Garz\'on, J. 2010, accepted in ApJ
  (arXiv1009.0112R)

\bibitem[{Ulvestad \& Wilson(1989)}]{B2}
Ulvestad, J.~S., \& Wilson, A.~S. 1989, ApJ, 343, 659

\end{thebibliography}

\end{document}